\documentclass[useAMS,usenatbib]{mn2e}

\usepackage{graphicx}
\usepackage{times}
\usepackage{natbib}

\begin{document}
 
\title[Constraining the electron and proton acceleration efficiencies in merger shocks in galaxy clusters ]{Electron and proton acceleration efficiency by merger shocks in galaxy clusters }
\author[F. Vazza, D. Eckert, M. Br\"{u}ggen, B. Huber]{F. Vazza$^{1}$\thanks{%
 E-mail: franco.vazza@hs.uni-hamburg.de}, D. Eckert$^{2}$, M. Br\"{u}ggen$^{1}$, B. Huber$^{3}$\\
$^{1}$ Hamburger Sternwarte, Gojenbergsweg 112, 21029 Hamburg, Germany \\
$^{2}$ Astronomy Department, University of Geneva 16, ch. d'Ecogia, CH-1290 Versoix Switzerland\\
$^{3}$ Royal Institute of Technology (KTH), SE-106 91 Stockholm, Sweden}
\date{Accepted ???. Received ???; in original form ???}
\maketitle

\begin{abstract}
Radio relics in galaxy clusters are associated with powerful shocks that (re)accelerate relativistic electrons. It is widely believed that the acceleration proceeds via diffusive shock acceleration. In the framework of thermal leakage, the ratio of the energy in  relativistic electrons to the energy in relativistic protons should should be smaller than $K_{\rm e/p} \sim 10^{-2}$. The relativistic protons interact with the thermal gas to produce $\gamma$-rays in hadronic interactions. Combining  observations of radio relics with upper limits from $\gamma$-ray observatories can constrain the ratio $K_{\rm e/p}$. 
In this work we selected 10 galaxy clusters that contain double radio relics, and derive new upper limits from the
stacking of $\gamma$-ray observations by \emph{Fermi}. We modelled the propagation of shocks using a semi-analytical model, where we assumed a simple geometry for shocks and that cosmic ray protons are trapped in the intracluster medium.  Our analysis shows that diffusive shock acceleration has difficulties in matching simultaneously the observed radio emission and the constraints imposed by \emph{Fermi}, unless the magnetic field in relics is unrealistically large ($\gg 10 ~\rm \mu G$). In all investigated cases (also including realistic variations of our basic model and the effect of re-acceleration) the mean emission of the sample is of the order of the stacking limit by  \emph{Fermi}, or larger. These findings put tension on the commonly adopted model for the powering of radio relics, and imply that the relative acceleration efficiency of electrons and protons is at odds with predictions of diffusive shock acceleration, requiring $K_{\rm e/p} \geq 10-10^{-2}$. 
\end{abstract}

\label{firstpage}
\begin{keywords}
Galaxy clusters; intergalactic medium; shock waves; acceleration of particles; gamma-rays.
\end{keywords}

\section{Introduction}
\label{sec:intro}

Radio relics are steep-spectrum radio sources that are usually detected in the 
outer parts of galaxy clusters, at distances of $\sim 0.5-3$ Mpc from their centres.  They are very often found in clusters with a perturbed dynamical 
state. There is good evidence for their association with powerful merger shocks, as early suggested by \citet[][]{1998A&A...332..395E}. 
Among them are {\it giant double-relics} that show two large sources on opposite sides of the host cluster's centre \citep[e.g., ][]{fe12,2012MNRAS.426...40B,fdg14}. These relics are associated with shocks in the intracluster medium that occur in the course of cluster mergers.  Only a tiny fraction of the kinetic power dissipated by typical cluster merger shocks ($\ll 10^{-3}$) is necessary to power the relics, and diffusive shock acceleration (DSA, \citealt[e.g.][for modern reviews]{2012JCAP...07..038C,kr13}) has so far been singled out as the most likely mechanism to produce the relativistic electrons and to produce the observed power-law radio spectra  \citep[e.g.][]{hb07}.
However, if the standard DSA model is correct, the same process should also lead to the acceleration of cosmic-ray (CR) protons. Indeed, the process should be much more efficient for protons, owing to their larger Larmor radius {\footnote{Instead, since the Larmor radius of thermal electrons is much smaller than the typical shock thickness, thermal electrons {\it cannot} be easily accelerated to relativistic energies by DSA. This so called injection problem for electrons is still largely unresolved \citep[e.g.][]{2014IJMPD..2330007B}}}. \\
To date, high-energy observations of nearby galaxy clusters have not revealed any diffuse $\gamma$-ray emission resulting from the interaction between relativistic protons and thermal particles of the intracluster medium (ICM) \citep[][]{re03,aha09,al10,alek12,arl12,2014MNRAS.440..663Z}. Recently, the non-detection of diffuse $\gamma$-ray emission from clusters by \emph{Fermi} has put the lowest upper limits on the density of CRs in the ICM, $\leq$ a few percent of the thermal gas energy within the clusters virial radius \citep{ack10,fermi14}. 
Moreover, the stacking of subsets of cluster observations leads to even lower upper-limits \citep[][]{2013A&A...560A..64H,2014ApJ...795L..21G}.
These low limits on the energy content of CRs can be used to constrain shock-acceleration models.
Recently, \citet{va14relics} have already suggested that the present statistics of radio observations, combined with available upper limits by \emph{Fermi} places constraints on DSA as the source of giant radio relics. In \citet{va14relics}, we assumed that the population of clusters with radio relics was similar to the population of (non-cool-core) clusters for which the stacking of \emph{Fermi} clusters was available. In the present paper, we repeat a similar analysis by comparing to a more realistic stacking of the 
\emph{Fermi} data. Our method is outlined in Sec.~\ref{subsec:algo}, while our results are given in Sec. \ref{sec:results}. In the latter Section, we also discuss on the role 
played by the several open parameters in our modeling. We find (Sec.~\ref{sec:results}) that the present upper limits from \emph{Fermi} imply energy densities of CR-protons that are too low to be explained by standard DSA: if 
DSA produces the electrons in relics, then we should have already detected hadronic $\gamma$-ray emission in some clusters, or in stacked samples.  In our conclusions (Sec.~\ref{sec:conclusions}), we discuss possible solutions to this problem, as suggested by recent hybrid and particle-in-cell simulations of weak, collisionless shocks. 

\section{Methods}
\label{sec:methods}

\subsection{Semi-analytical cluster mergers}
\label{subsec:algo}
Our aim is to test DSA by making  quantitative predictions of the hadronic $\gamma$-ray flux assuming that the CR-protons come from the same shocks that produce radio relics (Sec.~\ref{subsubsec:radio}). Semi-analytical methods with initial conditions tuned to match observable parameters of radio halos have been widely used in the literature \citep[e.g.][]{2003ApJ...583..695G,2005MNRAS.357.1313C,hb07}. They have obvious limitations owing to the lack 
of 3-dimensional detail and the crude geometrical assumption on the merger scenario (i.e. spherical symmetry and simple radial distribution for the gas). Still, for this specific problem they are helpful, as the energy density of CR-protons should be a simple function of the shock parameters and of the volume crossed by each shock wave.\\

\begin{figure}
    \includegraphics[width=0.49\textwidth]{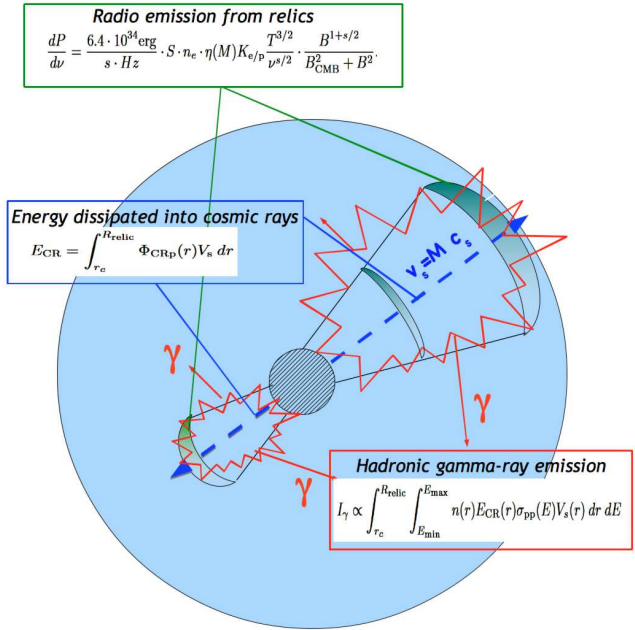}   
     \caption{Schematic view of our method for computing $\gamma$-ray emission from accelerated CR-protons downstream of the double relics.}
   \label{fig:sketch}
 \end{figure}

Our approach is similar to the method used in \citet[][]{va14relics}: for each cluster in our sample, we model the shock trajectories and the associated CR-acceleration.  We use simple 1D models of shock propagation in stratified atmospheres and use observables from radio and X-ray data, such as the spectral index, the distance from the centre, the radio power and the largest linear scale of each relic \citet{2012MNRAS.426...40B} (see also \citealt{fdg14}).
Below we summarise the most important steps from \citet[][]{va14relics} (see also Fig.~\ref{fig:sketch}), while in Sec.\ref{subsec:uncert} we discuss the most relevant uncertainties in our model and assess their impact on our
results:

\begin{enumerate}
\item We infer the Mach number, $M$, from the spectral index of the radio spectrum, $s$, at the injection region via $M= \sqrt \frac{\delta+2}{\delta-2}$ ($\delta$ is the slope of the particle energy spectra and $\delta=2 s$). This assumes that the radio spectrum is dominated by the freshly injected CR-electrons at the shock, an assumption that might be poor for spectra integrated over large downstream volumes (in which cases the radio spectrum is steeper by $0.5)$. Bootstrapping with random deviates from the Mach number thus determined can quantify the errors (see Sec.~\ref{sec:results}). 

\item The upstream (i.e. pre-shock) gas density, $n_{\rm u}$, at the relic is computed  using a $\beta$-model profile for each host cluster, with $\beta=0.75$ and the core radii scaling as $r_{\rm c}=r_{\rm c, Coma}(T_{\rm Coma}/T)^{1/2}$ (which follows from the self-similar scaling), where $r_{\rm c, Coma}=290$ kpc. This is the only way to regularise our dataset, as a more detailed reconstruction of the gas density from the literature is missing for most of these objects. Both are clearly a simplification but in reality we probably find higher gas densities and temperatures along the merger axis, due to clumping and enhanced gas compression which will only increase the $\gamma$-ray emission. Moreover, the propagation of each merger shock is determined by the temperature {\it in front of it}, i.e. in the upstream region, which present X-ray observation can hardly constrain for any of these objects. Therefore  we always consider an upstream gas temperature, $T_{\rm u}$, based on the $L_{\rm X} - T_{\rm 500}$ relation for each host cluster \citep[][]{2009A&A...498..361P}. 

\item  We compute the  kinetic power for each shock as $\Phi_{\rm kin}=n_{\rm u} v_{\rm s}^3  S/2 $, where $v_{\rm s}=M c_{\rm s}$ ($c_s \propto \sqrt{T_{\rm u}}$).

\item We assume that a fraction of the kinetic power goes into CR-protons $\Phi_{\rm CRp}=\eta(M)\Phi_{\rm kin}$, where the efficiency, $\eta(M)$, is a non-linear function of the Mach number. It has been derived for several DSA scenarios \citep[e.g.][]{kj07,kr13,2014ApJ...785..133H}. Here we use $\eta(M)$ given by \citet{kr13}, which were estimated based on simulations of nonlinear DSA,
considering an upstream $\beta=100$ plasma and including a phenomenological model for the magnetic field amplification and Alfvenic drift
in the shock precursor, due to accelerated CRs. This function predicts an acceleration efficiency of $\approx 1$ percent for $M=3$, steeply rising to $\sim 10$ percent for $M=5$ \citep[see Fig.~4 of ][]{kr13}.  The corresponding power into CR-electrons is set by assuming a fixed electron-to-proton ratio, $K_{\rm e/p}=0.01$: $\Phi_{\rm CRe}=K_{\rm e/p}\eta(M)\Phi_{\rm kin}$.  This ratio is already conservative, as recent models of particle acceleration in supernovae suggest an even lower value, $K_{\rm e/p} \sim 10^{-3}$ \citep[][]{2014arXiv1412.0672P}, which would result into a $10$ times larger hadronic $\gamma$-ray flux than for our value of $K_{\rm e/p}$.

\item The magnetic field at the relic, $B$, is derived from the radio power via the equations given by \citet{hb07}. In this model
the monochromatic radio power at frequency $\nu$, $P_{\nu}$, depends on the shock surface area, $S$ (which we derive from the projected size of the relic, assuming that the relic has a circular shape), the upstream electron density, $n_{\rm e}$ (computed from $n_{\rm u}$ by assuming a mean molecular mass of $0.6$), the upstream electron temperature,  $T_{\rm e} \approx T_{\rm u}$,  the spectral index of the radio emission, $s$, and the relic magnetic field, $B_{\rm relic}$, in the following way:

\begin{equation}
\frac{dP}{d\nu}=\frac{6.4 \cdot 10^{34} \rm erg}{\rm s \cdot Hz} \cdot S \cdot n_e \cdot  \eta(M) K_{\rm e/p} \frac{T_{\rm e}^{3/2}}{\nu^{s/2}} \cdot \frac {B_{\rm relic}^{1+s/2}}{B_{\rm CMB}^2+B_{\rm relic}^2} ,
\label{eq:hb}
\end{equation}
where $B_{\rm CMB}$ is the equivalent field of the cosmic microwave background.{\footnote {We notice that, compared to the original Equation by \citet{hb07}, we use here upstream values for density and temperature, as the function $\eta(M)$ also accounts for the shock compression factors.}} In our simplest model ("basic model") we let the magnetic field
as a free parameter without upper limits, while in a more realistic scenario ("Bcap" scenario) we imposed a maximum magnetic field of $B_{\rm relic,max}=10 ~\mu G$ for all relics (Sec.\ref{subsubsec:Bfield}).

\item Explaining the observed radio emission from $M \leq 3$ shocks is a problem, as the required electron acceleration efficiency at these weak shocks can become unrealistically large \citep[][]{2011ApJ...728...82M}. It has been suggested that the contribution from shock re-accelerated electrons can alleviate this problem, as it would mimic the effect
of having a higher acceleration efficiency  \citep[][]{ka12,pinzke13}. To model re-acceleration, as in \citet[][]{va14relics},
we used an increased CR-proton acceleration efficiency as a function of Mach number, following \citet[][]{ka07} and \citet{kr13} who showed that the net effect of re-accelerated and freshly injected cosmic rays can be modeled by a rescaled efficiency $\eta(M)$. This depends on the
energy ratio between pre-existing cosmic rays ($E_{\rm CR}$) and the thermal gas ($E_{\rm g}$). In the following we parametrize this ratio using the parameter $\epsilon=E_{\rm CR}/E_{\rm g}$, and explore the cases $\epsilon=0$ (single injection, no-reaccelerated electrons), and bracket the trend of re-acceleration using $\epsilon=0.01$ and  $\epsilon=0.05$. Notice that the latest limits from \emph{Fermi} only allows $\epsilon \leq $ a few percent, for flat radial distributions of CRs \citep[][]{fermi14}.

\item In order to compute the spectrum of electrons in the case of a re-accelerated pool of CRs, we follow \citet[][]{ka12} and
assume that the blend of several populations of pre-existing CR-electrons are characterised by a power law with index $\delta_e = (4\delta+1)/3$, where $\delta$ is the spectral index of the energy spectrum derived at the relic as before. As in \citet[][]{ka12} we fix the spectrum of re-accelerated CRs to the particle spectrum at the relic in those cases where the spectrum is flatter than the (steep) spectrum associated with the $M=2$ shock that is supposed to re-accelerate them. 

\item Once the shock parameters are fixed, we estimate the energy injected in CR-protons that are assumed to stay where they have been predicted. We assume that each shock surface scales with the cluster-centric distance, $r$, as $S(r)=S_0 (r/R_{\rm relic})^2$ (i.e. the lateral extent of the shock surface is set to largest linear size of the relic, which decreases with $\propto r$ inwards). The cumulative energy dissipated into CR-protons is given by 

\begin{equation}
E_{\rm CR}= \int_{r_c}^{R_{\rm relic}}{\Phi_{\rm CRp}(r)v_{\rm s} ~ dr } = \int_{r_c}^{R_{\rm relic}} \frac{\eta(M)  n_{\rm u}(r) v_{\rm s}^3  S(r)}{2}  ~ dr.
\end{equation}
The shock surface and strength will vary with radius, and so will the CR-proton acceleration efficiency, $\eta(M)$. Our fiducial model assumes that the Mach number of shocks released by the merger is constant across the whole volume of interest, while in 
Sec. \ref{subsubsec:Mrad} we also test a scenario in which the Mach number scales with radius. The lower integration limit in the equation for $E_{\rm CR}$ is the core radius, $r_c$, meaning that the shocks are assumed to be launched only outside of the cluster core, which is supported by simulations \citep[e.g.][]{va12relic,sk13}

\item We compute the hadronic $\gamma$-ray emission, $I_{\rm \gamma}$, following \citep[][]{pe04,donn10},with the only difference that for the hadronic cross-section we use the parametrisation of the proton-proton cross section given by \citet{2006PhRvD..74c4018K}, as in  \citet{2013A&A...560A..64H}. 
In detail, we compute for each radius the source function of $\gamma$-rays as:
\begin{eqnarray}
q_{\gamma}(E_{\gamma})={{ 2^{4-\delta_{\gamma}} }\over{
3\delta_{\gamma}}}
{{\sigma_{\mathrm{pp}}(E) c n_{\mathrm{e}}(r) K_p}\over{\delta -1}}
(E_{\mathrm{min}})^{-\delta} \frac{E_{\mathrm{min}}}{\mathrm{GeV}} \times\frac{m_{\pi^{0}}c^{2}}{\mathrm{GeV}},
\end{eqnarray}

where $n_{\rm e}$ is the upstream electron density, computed from the gas density by assuming a molecular mean weight $\mu=0.6$. The spectrum of the $\gamma$-ray emission depends on the assumed Mach number across the cluster, and therefore  is either a function of $M(r)$ in the single acceleration model, or a constant in the re-acceleration case. Once the spectral index, $\delta$, of the particle spectrum is fixed, the spectrum of the $\gamma$-ray emission is given by $\delta_{\gamma}=4(\delta-1/2)/3$ \citep[][]{pe04}. Here, we consider hadronic emission in the energy range [0.2-300] GeV, which are compared to the stacked emission from \emph{Fermi} described in the following section (Sec.~\ref{subsubsec:gamma_obs}). $m_{\pi}$ and $m_p$ are the masses of the $\pi^0$ and the proton, respectively.  The threshold proton energy is taken as $E_{\rm min}=780 ~\rm MeV$ and the maximum is  $E_{\rm max}=10^{5}$ MeV (the actual value is actually irrelevant, given the steep spectra of our objects)
The  effective cross-section we used is

\begin{equation}\label{eq:sigma}
\sigma_{pp}(E)=(34.3+1.88L+0.25L^2)\left[1-\left(\frac{E_{th}}{E}\right)^2\right]^4 \mbox{mb},
\end{equation}
with $L=\ln(E/\mbox{1 TeV})$ and $E_{th}=m_p+2m_{\pi}+m^2_{\pi}/2m_p\sim1.22$ GeV \citep[Eq. 79 of ][]{2006PhRvD..74c4018K}. The normalisation factor $K_p$ is: 
\begin{equation}
K_p=\frac{(2-\delta)E_{\rm CR}(r)}{(E_{\rm max}^{-\delta+2}-E_{\rm min}^{-\delta+2})}.
\end{equation}

The emission per unit of volume, $\lambda_{\gamma}$, is
obtained by integrating the source function over the energy range: 

\begin{eqnarray}
\label{eq:gamma}
        \lambda_{\gamma}(r) &= \int\limits_{E_{1}}^{E_{2}}
        \mathrm{d}E_{\gamma} q_{\gamma}(E_{\gamma}) \nonumber \\ &=
        \frac{\sigma_{\mathrm{pp}} m_{\pi}c^{3}}{3
        \delta\delta_{\gamma}}
        \frac{n_{\mathrm{u}}(r) K_p}{\delta -1}
        \frac{ (E_{\mathrm{min}})^{-\delta}}{2^{\delta_{\gamma}-1}}
        \frac{E_{\mathrm{min}}}{\mathrm{GeV}}  \left(
        \frac{m_{\pi_{0}}c^{2}}{\mathrm{GeV}}
        \right)^{-\delta_{\gamma}} \nonumber \\ &\times
        \left[\mathcal{B}_{\mathrm{x}}\left(
          \frac{\delta_{\gamma}+1}{2\delta_{\gamma}},
          \frac{\delta_{\gamma}-1}{2 \delta_{\gamma}} \right)
          \right]^{x_{1}}_{x_{2}}
\end{eqnarray}
where $\mathcal{B}_{\mathrm{x}}(a,b)$ denotes the incomplete
$\beta$-function and $[f(x)]^{a}_{b} = f(a) - f(b)$. Integrations over radii are performed by summing over radial shells of thickness 10 kpc. 
Finally, the hadronic $\gamma$-ray emission for each radial shell along the trajectory of the shock is given by $\lambda_{\gamma}(r) S(r)  ~dr$ and the total hadronic emission in the downstream region of each relic is given by the integral, $I_{\rm \gamma}=  \int_{r_c}^{R_{\rm relic}} \lambda_{\gamma}(r) S(r) ~dr$.
\end{enumerate}

The above set of approximations minimises the hadronic $\gamma$-ray emission from clusters:  the presence of gas clumping and substructure is neglected, the injection of CRs is limited to regions outside of the dense
cluster cores, and additional acceleration of CRs by earlier shocks, turbulence, supernovae and AGN is not taken into account. A number of assumptions have to be made in the previous steps. We discuss all most important in Sec. \ref{subsec:uncert} and explore the effects uncertainties in the parameters of the model.

\subsection{Observations}
\label{subsec:obs}

\subsubsection{Radio data}
\label{subsubsec:radio}

We restrict our analysis of to double radio relics, as these systems are caused most clearly by major merger events \citep[e.g.][]{1999ApJ...518..594R,vw12sim,sk13}, and they should be less affected by projection effects because of large cluster-centric distances \citep[][]{va12relic}. We select double-relic sources from the collection of \citet{2012MNRAS.426...40B} and further restrict analysis to the sources at high Galactic latitude ($|b|>15^{\circ}$) to avoid strong contamination by the bright diffuse $\gamma$-ray emission from the Galactic plane (see below).  The final sample is made of 20 relics from 10 clusters: MACSJ1752, A3667, A3376, A1240, A2345, A3365, MACSJ1149, MACSJ1752, PLCKG287, ZwClJ2341 and RXCJ1314. The values of radio parameters for these objects (e.g. total power, radio spectral slope, $s$, largest linear scale
of each object and distance from the centre of the host cluster) are given in Tab.1.

\subsubsection{$\gamma$-ray data}
\label{subsubsec:gamma_obs}

We analysed $\gamma$-ray data collected by the Large Area Telescope (LAT) on board \emph{Fermi} (hereafter \emph{Fermi}-LAT). \emph{Fermi}-LAT \citep{atwood09} is a pair-conversion $\gamma$-ray telescope operating in the $20$ MeV - $300$ GeV band. We collected all the data obtained during the period 2008-08-04 to 2013-11-01 and analysed them using the \emph{Fermi} Science Tools software package \texttt{v9r32p5} and the \texttt{P7SOURCE\_V6} instrument response files. For each source listed in Table 1, we extracted a rectangular region of interest (ROI) with side $\sim 20^{\circ}\times 20^{\circ}$ centred on the source. For each source, we then constructed a model including the Galactic emission \footnote{http://fermi.gsfc.nasa.gov/ssc/data/analysis/software/aux/gal\_2yearp7v6\_v0.fits}, the isotropic diffuse background \footnote{http://fermi.gsfc.nasa.gov/ssc/data/analysis/software/aux/iso\_p7v6source.txt}
 and all sources listed in the 2FGL catalog in a radius of 20$^\circ$ around our source, with spectral models and parameters fixed to the values given in 2FGL. We then performed a binned likelihood analysis \citep{mattox96} on each individual ROI to determine the parameters of the $\gamma$-ray emission model. The normalization of each component (diffuse background components and point sources) was left free to vary during the fitting procedure. \\

As an alternative hypothesis, we then added to the model a pointlike test source fixed to the cluster position. The test source had a spectrum computed from the simulated CR proton population, accelerated at shocks (Sec. \ref{subsec:algo}), which should be injected by $M \sim 3$ shocks in most cases. This spectrum is the best guess from our cosmological numerical simulations (\citealt{scienzo14}) and is close to the "universal CR-spectrum" suggested by \citet[][]{2010MNRAS.409..449P}.
The resulting photon spectrum was discretised and used as a template for the expected emission of the shock-accelerated CR protons. This spectral model closely resembles a power law with a photon index of $\sim2.6$ at energies $>1$ GeV and is significantly flatter at lower energies because of the strong dependence of the $p-p$ interaction cross section near the threshold energy for pion production \citep[see Appendix A.1 of ][]{2013A&A...560A..64H}. We then estimated the test statistic (TS) defined as 
\begin{equation}
\mathrm{TS} = -2\,\mathrm{ln}\, \frac{\mathcal{L}^{max}_{0}}{\mathcal{L}^{max}_{1}},
\label{eq:TS}
\end{equation}
where $\mathcal{L}^{max}_{0}$ and $\mathcal{L}^{max}_{1}$ are the maximum likelihood values for the null and alternative hypotheses, respectively. No significant signal ($TS>25$) was observed for any of the clusters from Table 1. Therefore, we computed upper limits at the 95\% confidence level (CL) for all sources as in \citet{fermi14}, which we report in Table 3 for two energy ranges ($0.2-300$ and $1-300$ GeV). The choice of a pointlike source at the cluster positions is not critical. The mean spectra assumed for the relics are steep, and most of the photons should come from the low energy end of the spectrum, where the resolution of \emph{Fermi} is too coarse to distinguish between poinlike and extended sources in our objects ($\sim 4^{\circ}$ at $200$ MeV).  \\

\begin{figure}
\includegraphics[width=0.45\textwidth]{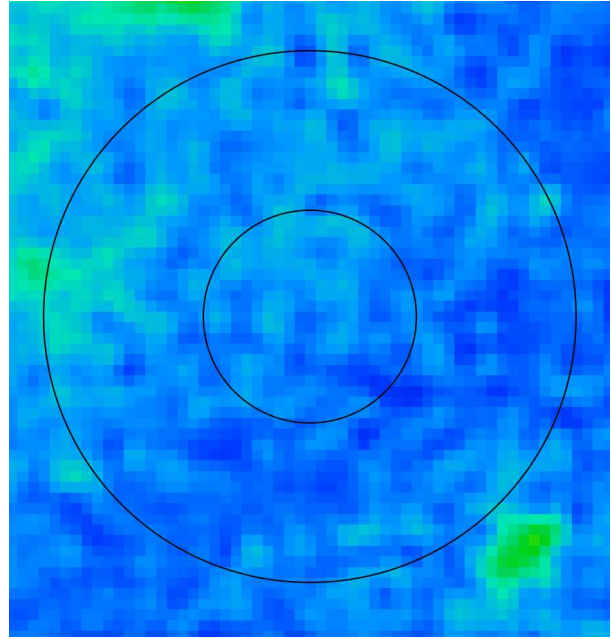}   
\caption{Co-added \emph{Fermi}-LAT count map in the 200 MeV - 300 GeV energy range for all the clusters listed in Table \ref{tab:tab1}. The circles indicate apertures of 2.5$^\circ$ and 5$^\circ$ around the stacked cluster position to highlight the source and background regions, respectively.}
   \label{fig:stack_fermi}
 \end{figure}

In order to lower these limits, we performed a stacking analysis of the sources in our sample following the method presented in \citet{huber12,2013A&A...560A..64H}. The stacking of the sources is performed by adding step by step the individual ROIs after having simulated and subtracted the surrounding point sources. The resulting co-added map for our stacked sample is given in Fig.~\ref{fig:stack_fermi}. The co-added data are then fit using the same source+background model as described above. Again, no significant emission is measured within the stacked volume of these clusters ($TS_{\rm stacked}=0.21$), which results in an upper limit of $F_{\rm mean}<1.90\cdot 10^{-10}$ \rm ph/cm$^{2}$/s (95\% CL, $0.2-300$ GeV band) to the average flux per cluster. For more details on the stacking procedure and a thorough validation of the method using simulated data, we refer the reader to \citet{huber12}.

\begin{table*}
\label{tab:tab1}
\caption{Main observational parameters for the radio relics and clusters considered in this paper: redshift (2nd column), X-ray luminosity (3rd), distance from the cluster centre for each relic in the pair 4th and 5th column), largest linear scale (6-7), radio power (8-9), radio spectral index (10-11) of relics, upper limits of $\gamma$-ray emission at $0.2-300$ and $1-300$ GeV for each host cluster (12-13) and value of the test statistics (Eq. \ref{eq:TS}). The radio data are taken from \citet{fe12} and \citet{2012MNRAS.426...40B}, while the $\gamma$-ray data have been derived from the \emph{Fermi} catalogue in this work.}
\centering \tabcolsep 5pt 
\begin{tabular}{c|c|c|c|c|c|c|c|c|c|c|c|c|c|c|c|c}
object & z & $L_x$ & $r_1$ & $r_2$ & $R_1$ & $R_2$ & $\log_{\rm 10}(P_{\rm R,1})$ & $\log_{\rm 10}(P_{\rm R,1})$ & $s_1$ & $s_2$  & $\log_{\rm 10}(UL_{\rm 0.2-300})$ & $\log_{\rm 10}(UL_{\rm 1-300}$) & TS \\
  &  &  $10^{44}$ erg/s & Mpc & Mpc & Mpc & Mpc & [erg/s] & [erg/s] & & &   $[\rm ph/(s  ~cm^2)]$ & $[\rm ph/(s  ~cm^2)]$ & \\ \hline 
A3376 & 0.047 & 1.04 & 0.80 & 0.95 & 1.43 & 0.52 & 40.026 & 39.936 & 1.20 & 1.20   & -8.65 & -9.71 & 1.36 \\
A3365 & 0.093 & 0.41 & 0.56 & 0.23 & 1.00 & 0.70 & 40.086 & 39.146 & 1.55 & 1.93& -8.61 & -9.66 & 6.17\\
A1240 & 0.159 & 0.48 & 0.64 & 1.25 & 0.70 & 1.10 & 39.748 & 39.991 & 1.20 & 1.30  & -9.21 & -9.89 & 0\\
A2345 & 0.176 & 2.56 & 1.50 & 1.15 & 0.89 & 1.00 & 40.544 & 40.577 & 1.30 & 1.50 & -8.55 & -9.55 & 6.19\\
RXCJ1314 & 0.244 & 5.26 & 0.91 & 0.91 & 0.57 & 0.94 & 40.350 & 40.714 & 1.40 & 1.40  & -9.06 & -9.94 & 0\\
MACSJ1149 & 0.540 & 6.75 & 0.82 & 0.76 & 1.39 & 1.14 & 40.879 & 41.100 & 1.20 & 1.42  & -9.53 & -10.11 & 0\\
MACSJ1752 & 0.366 & 3.95 & 1.34 & 0.86 & 1.13 & 0.91 & 41.651 & 41.292 & 1.21 & 1.13 & -8.92 & -9.98 & 0.04\\
A3667 & 0.046 & 1.02 & 1.42 & 1.83 & 1.30 & 1.30 & 40.044 & 39.945 & 1.39 & 1.31 &  -8.59 & -9.64 & 3.16\\
ZwCLJ2341 & 0.270 & 2.00 & 0.25 & 1.20 & 1.18 & 0.76 & 40.726 & 40.377 & 1.90 & 1.22 & -9.41 & -10.35 & 0\\
PLCKG287 & 0.390 & 8.29 & 1.93 & 1.62 & 1.58 & 3.00 & 41.401 & 41.322 & 1.26 & 1.54 & -8.81 & -9.68  & 0.49\\
\end{tabular}
\end{table*}

\begin{figure}
    \includegraphics[width=0.45\textwidth]{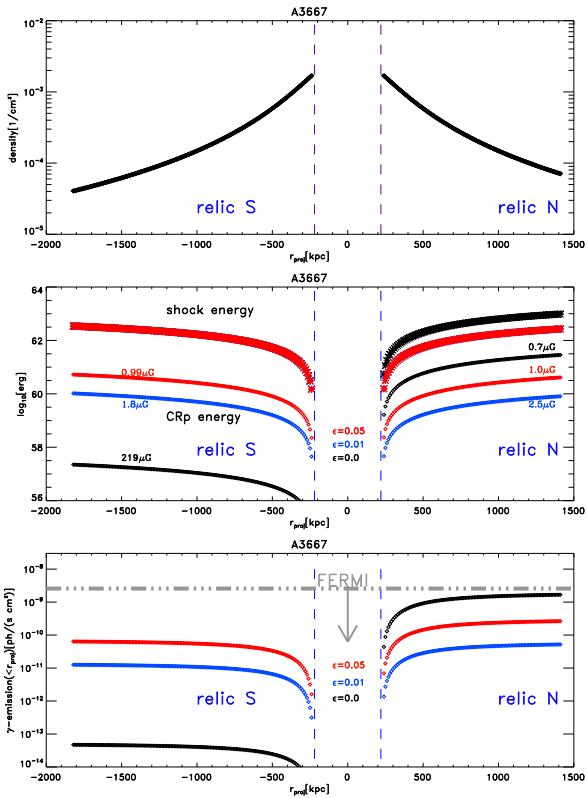}   
     \caption{1D profiles of various quantities along the projected propagation
radius of the two relics in A3667, as assumed or predicted by the basic model (Sec. \ref{subsec:algo}). First panel: gas density profile. Second panel: profile of the cumulative shock energy and CRp-energy dissipated in the downstream. The additional numbers in colours give the magnetic field necessary to reproduce the observed radio power using Eq. \ref{eq:hb}. Third panel: predicted $\gamma$-ray emission. The vertical lines delimit the regions where the shocks have been launched, while the horizontal line gives the single-object limit from \emph{Fermi}.}
  \label{fig:example}
\end{figure}

\section{Results}
\label{sec:results}

\subsection{Basic model}

A typical set of predictions from our baseline model is given in Fig.~\ref{fig:example} for the two relics in A3667.
The integrated kinetic energy that crosses the shock (upper lines) and dissipated CR-proton energy (lower lines) in the second panel are integrated along the shock trajectory. In the single injection model ($\epsilon=0$) the predicted acceleration of CRs is very different for the two relics, and follows from the 
different assumed Mach numbers: $M=3.8$ for the northern relic and $M=1.7$ for the southern relic. In the first case, the radio power is matched with the modest field strength of $B_{\rm relic} \approx 0.7 ~\rm \mu G$, while in the weaker southern shock the required magnetic field is $\approx 219\rm  ~\mu G$. Re-accelerated electrons
can explain the observed radio power in both cases using fields in the range $B_{\rm relic} \sim ~ 1-2 \rm \mu G$.  The $\gamma$-ray emission (lower panel) is the integrated hadronic emission in the downstream. Hence, the last bins on the left and on the right give the total emission from the two downstream regions and should be compared with the \emph{Fermi} limit for A3667. When the contribution from both relics is summed up, the predicted hadronic emission is very close to the {\it single-object} limit from \emph{Fermi} for this cluster. We find similar results for the relics in ZwCLJ2341 (see Tab. 2).
In the next sections we will discuss the predicted magnetic fields and the $\gamma$-ray emission from the full dataset and under different assumptions in our model. \\

\begin{figure*}
    \includegraphics[width=0.45\textwidth]{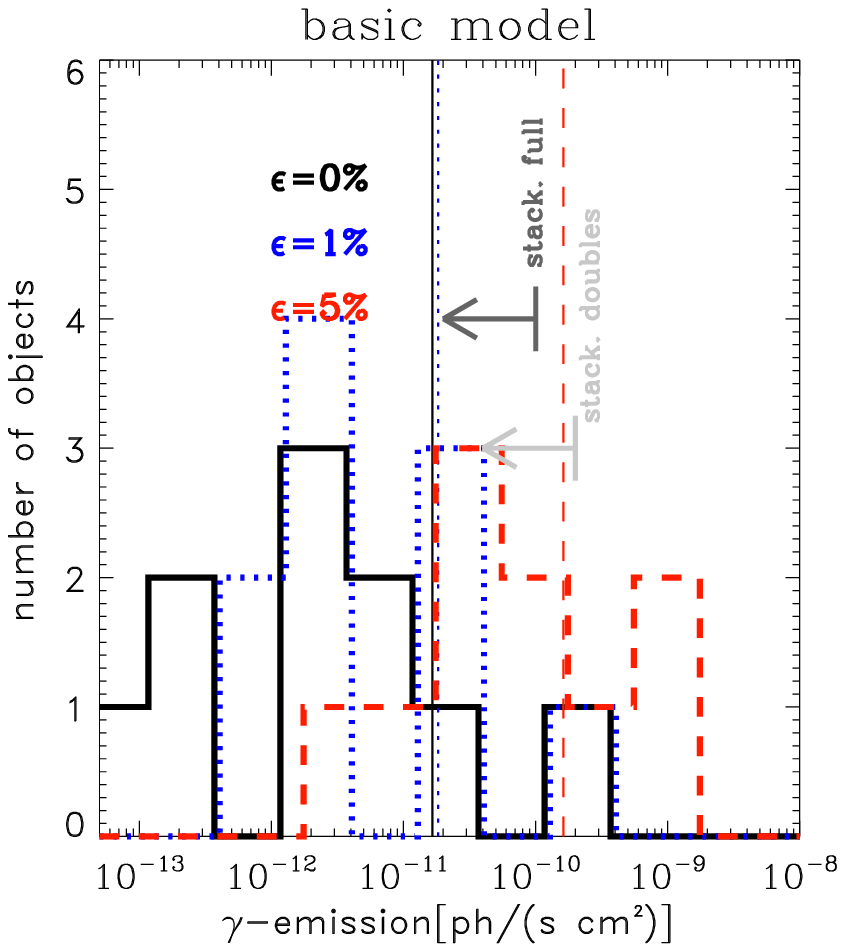}   
     \includegraphics[width=0.45\textwidth]{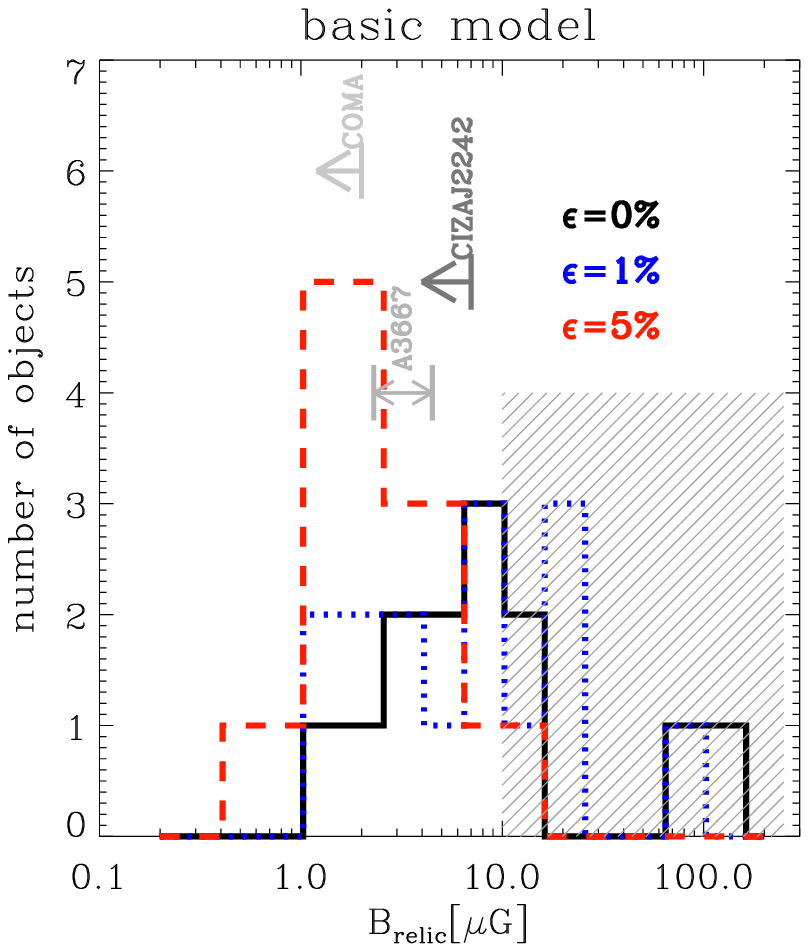}   
     \caption{Left: distribution of predicted $\gamma$-ray emission from our cluster sample, for different fiducial models (colored histograms). The thin vertical lines show the mean emission for the sample according to each model, and should be compared with the upper limits from the stacking of all non-cool-core clusters observed by \emph{Fermi}, or by the stacking limited to our sample of clusters with double relics (gray arrows). Right: distribution of magnetic fields (colored histograms) required for each relic to match the observed radio power using Eq. \ref{eq:hb}. The additional gray arrows give the range of values 
     inferred for the few observations of magnetic fields in real relics (see text for details), while the hatching marks the values of magnetic fields that we regard as physically unrealistic  (see Sec.\ref{subsubsec:Bfield}).}
       \label{fig:first}
  \end{figure*}

The full range of estimates from our fiducial model (Sec. \ref{subsec:algo}) is shown in Figure \ref{fig:first}, where we plot: a) the distribution of predicted $\gamma$-ray emission for the full sample (histograms with different colours for each model) and compare the mean emission of each run (thin lines) with the limits we derive from our stacking of  \emph{Fermi} exposures on these objects (Sec. \ref{subsubsec:gamma_obs}).  This stacking gives us the most robust testing of DSA, since it comes from the same set of objects simulated with our semi-analytical method. In the same figure, we also show for completeness the result of stacking only non-cool-core (NCC) clusters in a larger sample of objects observed by \emph{Fermi} \citep{2013A&A...560A..64H}, which we converted into the $[0.2-300]$ GeV energy range by assuming a $\gamma$-ray spectral index of $-2.6$. The limit here is $\sim 5$ times lower than our stacking limit (see also \citealt{2014ApJ...795L..21G} for a slightly lower limit), given the larger sample of objects (32) and the fact that this sample contains more nearby objects than our list of radio relics. This limit comes from a bigger population, yet comparing it to our simulated population is useful, under the hypothesis that two parent population of objects are dynamically similar. This is likely, because to the best of our knowledge all objects of our sample are NCC and all show evidence of very perturbed dynamical states in X-ray \citep[e.g.][see also http://www.mergingclustercollaboration.org/merging-clusters.html]{2003MNRAS.339..913E,cav09,2011ApJ...736L...8B,2012MNRAS.426...40B}. The magnetic fields at each relic required by our modelling of the radio power, using Eq.\ref{eq:hb}. We also show for comparison the upper limits derived on the magnetic field at the location of relics in the Coma cluster \citep[][based on the analysis of Faraday Rotation]{2013MNRAS.433.3208B} and in the cluster CIZA 2242.2+5301 \citep[][based on the analysis of the brightness profile across the relic]{vw10}, as well as the range of values inferred by \citet{2010ApJ...715.1143F}  for the relic north of A3667, based on the lack of Inverse Compton emission. \\

In our model, both $\epsilon=0$ and $\epsilon=5$\% runs predict a very high mean level of hadronic emission, well above the stacking of this dataset in the single injection case, and just below it in the case of $\epsilon=5$\% (but larger than the stacking of the full \emph{Fermi} catalog). In both cases the mean emission 
is kept at a high value by $1/5$ of bright objects in the sample (A3667 and ZwCLJ2341), however the bulk of all remaining objects is also characterised by emission of the the order of the full stacking of non-cool-core clusters. 
On the other hand, the $\epsilon=1$\% model predicts a mean emission below the the stacking of this dataset. However, the magnetic fields required in this case are very large. Here 6 out of 20 relics require  $B_{\rm relic} \geq 10  ~\rm \mu G$ where the radio emission is detected, which is at odds with the few available estimates of magnetic fields from observations (gray arrows). 
In a general sense, explaining fields larger than a few $\sim \rm \mu G$ at the large radii where these relics are found is difficult based on both observational and theoretical facts (see Sec.~\ref{subsubsec:Bfield} for a detailed discussion).\\
To summarise, none of the scenarios we investigated with our semi-analytical method is able to simultaneously predict a level of $\gamma$-ray emission compatible with \emph{Fermi} and to use a reasonable magnetic field level in all objects. The reacceleration model assuming $1$ \% of CR energy to be reaccelerated by merger shocks survives the comparison with \emph{Fermi}, but makes use of very large magnetic fields in $\sim 1/3$ of our objects. The models with single-injection or significant re-acceleration instead predict a mean emission for the sample which is in tension with stacking of \emph{Fermi} observations for this sample, or for the larger sample of non-cool-core clusters, which very likely has the same characteristics of the double-relics one.  

The following sections will show how this problem is worsened as soon as all relevant assumptions made in our modeling are relaxed.

\begin{table*}
\label{tab:tab_gamma}
\caption{Forecast of hadronic $\gamma$-ray emission from the downstream region of our simulated clusters in the $[0.2-300]$ GeV range. The first three columns show the prediction for the three assumed ratios $\epsilon$, and without imposing a maximum magnetic field at relics. The second three columns show our predictions by imposing a $B \leq 10 \mu G$ cap to the magnetic field. 
(see Sec. \ref{subsubsec:Bfield} for details). For comparison, we show the upper limits from \emph{Fermi} given in Table 1 for the same energy range. We mark with stars the objects for which the single-object comparison with {\emph Fermi} data is problematic in several models.}
\centering \tabcolsep 2pt 
\begin{tabular}{c|c|c|c|c|c|c|c|c|c|c}
model & $\epsilon=0$ & $\epsilon=0.01$ & $\epsilon=0.05$& $\epsilon=0, B_{\rm cap}$ &  $\epsilon=0.01, B_{\rm cap}$ &  $\epsilon=0.05, B_{\rm cap}$ & observed \\
 & $\log_{\rm 10}(e_{\gamma})$ & $\log_{\rm 10}(e_{\gamma})$ & $\log_{\rm 10}(e_{\gamma})$ & $\log_{\rm 10}(e_{\gamma})$ & $\log_{\rm 10}(e_{\gamma})$ & $\log_{\rm 10}(e_{\gamma})$  & $\log_{\rm 10}(UL_{\rm Fermi})$\\
object  &$[\rm ph/(s ~cm^2)]$ & $[\rm ph/(s ~cm^2)]$ & $[\rm ph/(s ~cm^2)]$ & $[\rm ph/(s ~cm^2)]$ & $[\rm ph/(s ~cm^2)]$ & $[\rm ph/(s ~cm^2)]$ &$[\rm ph/(s ~cm^2)]$ \\ \hline
A3376 &  -10.07 &  -10.54 & -9.83 & -9.78 & -9.77 & -9.77 & -8.65 \\
A3365 & -12.52 & -12.12 & 11.41 & -11.94 & -12.02 & -11.31 &  -8.61 \\
A1240 & -10.67 & -11.60 & -10.90 & -10.84 & -10.83 & -10.84 &-9.21 \\
A2345 &  -10.24 & -10.75 & -10.05 & -10.00 & -9.98 & -9.98 & -8.55\\
RXCJ1314 & -10.69 & -10.94 & -10.24 & -10.18 & -10.16 & -10.16 & -9.06 \\
MACSJ1149 & -11.31 & -12.24 & -11.53 & -11.47 & -11.46 & -11.46 & -9.53\\
MACSJ1752 & -10.14 & -10.45 & -10.86 &  -10.80 & -10.75 & -10.79 &  -8.92 \\
*A3667* &  -8.53 & -9.28 & -9.28 & -9.23 & -9.21 & -9.20 &  -8.59 \\
*ZwCLJ2341* &  -8.80 & -9.81 & -9.11 & -9.05 & -9.04 & -9.04 &  -9.41 \\
PLCKG287 & -10.75 & -11.44 & -10.74 & -10.68 & -10.67 & -10.67 & -8.81 \\

\end{tabular}
\end{table*}

\subsection{Model uncertainties}
\label{subsec:uncert}

\subsubsection{Magnetic field}
\label{subsubsec:Bfield}

\begin{figure*}
    \includegraphics[width=0.45\textwidth]{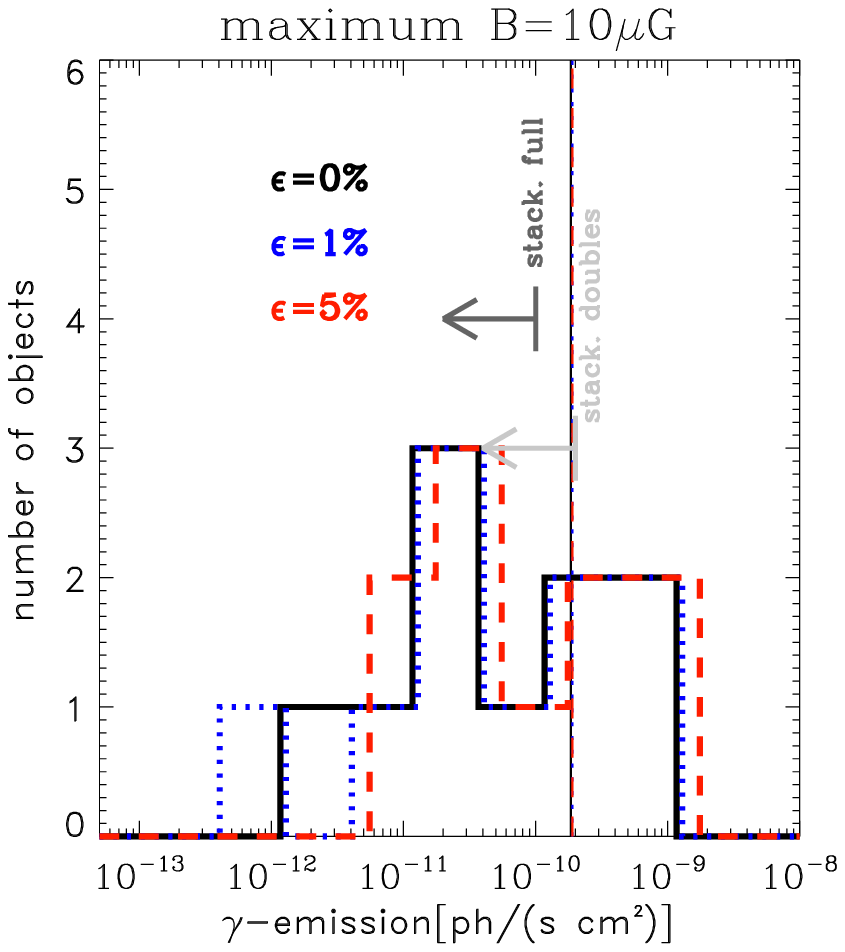}   
     \includegraphics[width=0.45\textwidth]{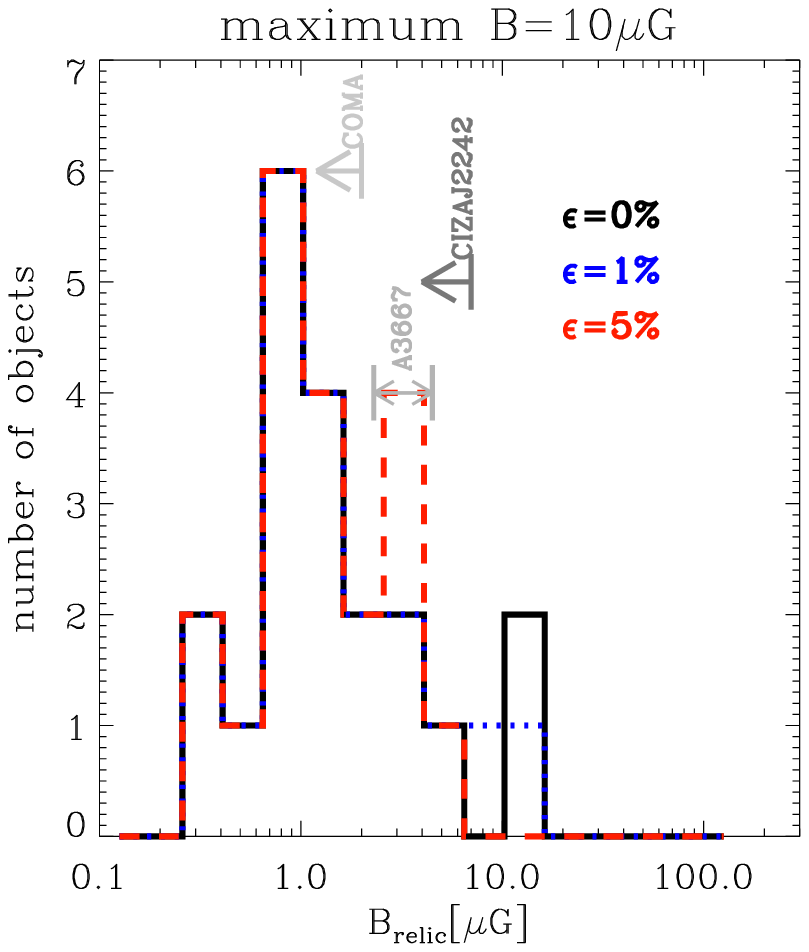}   
     \caption{Similar to Fig.\ref{fig:first}, but here we impose a maximum magnetic field of $10  ~\rm \mu G$ at the location of relics, and allow CR re-acceleration in all models (See Sec.~\ref{subsubsec:Bfield} for details).}
     \label{fig:Bcap}
  \end{figure*}

First, we investigate the role played by the magnetic field in relics. Explaining magnetic fields larger than a few $\rm \mu G$ outside of cluster cores is very difficult for several reasons.\\
In the only case in which a good volume coverage of the ICM is obtained through Faraday Rotation \citep[i.e. in the Coma cluster,][]{bo10,2013MNRAS.433.3208B}, the inferred trend of magnetic field is $B \sim n^{-\alpha_{\rm B}}$, with $\alpha_{\rm B} \sim 0.5-0.9$, which implies that on average the field drops below $1 \mu G$ at half of the virial radius. This scaling is supported by simulations \citep[][]{do99,bo10,va14mhd} and it implies that the magnetic field in
the ICM is not dynamically important, i.e. $\beta_{\rm pl} \sim 100$ everywhere (where $\beta_{\rm pl}=n k_{\rm B} T / P_{\rm B}$ and $P_{\rm B}$ is the magnetic pressure). Instead, a field of the order $\geq 10 \mu G$ at half of the virial radius or beyond implies $\beta \leq 1$, which is
hard to justify theoretically. Indeed, the turbulence around the relic should be modest and dominated
by compressive modes, and can only raise the magnetic field by a small factor \citep[][]{2012MNRAS.423.2781I,sk13}. It has been suggested that CRs can cause magnetic field amplification via CR-driven turbulent amplification \citep[][]{2013MNRAS.tmp.2295B}, but not to the extreme level required by our modelling. Moreover, all cosmological simulations predict some local amplification at shocks, but this is always smaller than the steep
increase of the gas thermal pressure, due to Rankine-Hugoniot jump conditions \citep[][]{do99,br05,xu09,va14mhd}. Based on simulations, the observed mass-radio Luminosity relation for double relics is better explained by assuming magnetic fields of the order of $\sim 2 ~\rm \mu G$ in most objects \citep[][]{fdg14}.\\
Evidence of {\it polarised} radio emissions from a few radio relics (including a few contained in our sample here) exclude the presence of large $\gg 10 ~\mu G$ fields distributed in scales below the radio beam (a few $\sim \rm ~kpc$), which would otherwise totally depolarise the emission \citep[][]{vw10,bonafede11,2012MNRAS.426...40B}.
The highest (indirect) indication of magnetic fields in a giant radio relic so far is of the order of $\sim 7  ~\rm \mu G$ \citep[][]{vw10}. Moreover, all observations of Faraday Rotation outside of cluster cores are consistent with magnetic fields of a few $\mu \rm G$ at most \citep[][]{mu04,gu08,bo10,vacca10}. 
Finally, we notice that the acceleration efficiencies usually assumed within DSA \citep[e.g.][]{kj07,hb07,kr13} are based on the assumption of shocks running in a high $\beta$ plasma.  In the case of high magnetisation or relics,  $\beta \ll 1$,  the physics of the intracluster plasma changes dramatically and the efficiencies from DSA are not applicable anymore.
Hence, we test a scenario in which we 
 cap the magnetic field at $B_{\rm relic, max}=10 ~\rm \mu G$.  In the many cases where the magnetic field inferred from Eq.\ref{eq:hb} would be larger than the $10   ~\rm \mu G$ upper limit, we allow for the presence of reaccelerated electrons and  iteratively increase $\epsilon$ (by $0.1$ \% at each iteration) so that the acceleration efficiency is increased. We stop the iterations when the radio emission from Eq.\ref{eq:hb} matches the radio power (within a $10$ \% tolerance). We then assume this ratio to be constant in the downstream region {\footnote {In this case, the values of $\epsilon=0$, $\epsilon=1$\% and $\epsilon=5$\% quoted in the labels actually refer to the initial assumed value for each cluster, while the final value depends on the iterations described here. However, in all cases the iterations stopped after reaching $\epsilon \sim$ a few percent.}}.
Fig. \ref{fig:Bcap} shows our results (see also Table 2). All magnetic fields are now more in line with the range of uncertainties given by the (scarce) observational data. 
In this case, the difference between all models is reduced, given that the energy ratio $\epsilon$ had to be increased in several objects also when the starting model is the single-injection one. The model producing
the largest emission is the one with $\epsilon=5$\%, but the difference with the two others is now limited to a few percent in the $\gamma$-ray flux. In this case, all models are now at the level of the observed stacking for this dataset, and a factor
$\sim 2$ above the full stacking of non-cool-core clusters in \emph{Fermi} \citep[][]{2013A&A...560A..64H}.
We think that this set of runs gives the most stringent test to the DSA model because it includes the effect of CR re-acceleration to
explain radio relics for modest magnetic field \citep[][]{ka12,pinzke13}, yet this idea fails when also the hadronic emission from accelerated CRs is taken into account. 
In the following we will discuss the remaining model uncertainties based on the "Bcap" model.

\subsubsection{Upstream gas density and temperature}
\begin{figure*}
    \includegraphics[width=0.45\textwidth]{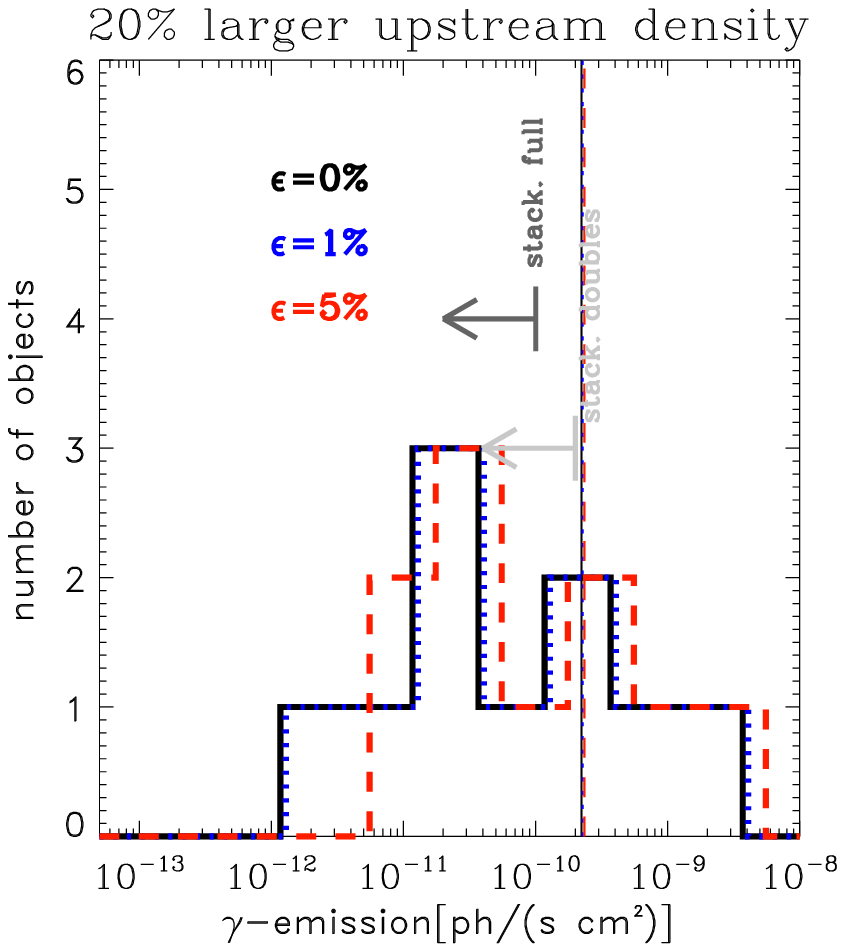}   
     \includegraphics[width=0.45\textwidth]{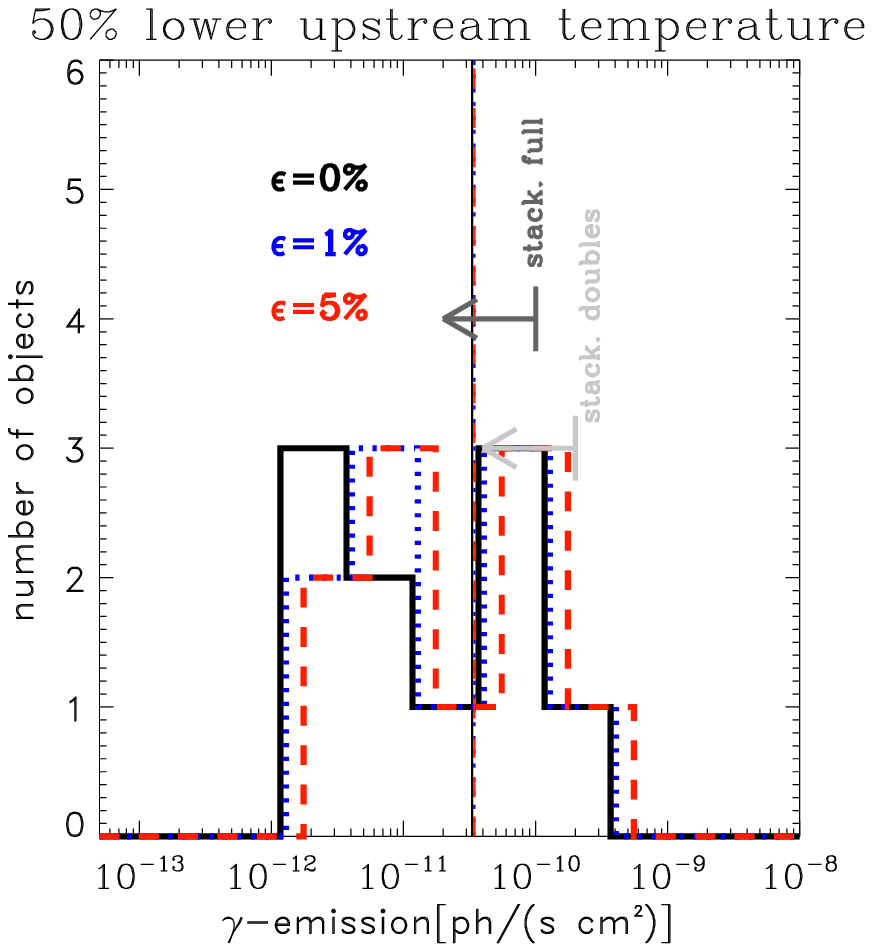}   
     \caption{Left panel: distribution of predicted $\gamma$-ray emission from our cluster sample, similar to Fig. \ref{fig:first} but here assuming a $20$ \% higher 
     upstream gas density (left) compared to the basic model. Right panel: as in the left panel but assuming a $50$ \% lower upstream gas temperature compared to the basic model (see text for details). In both cases we assume a capping of the magnetic field at $10 ~\rm \mu G$.}
       \label{fig:second}
  \end{figure*}

Our assumption for the upstream gas density follows from the simplistic assumption of a $\beta$-model profile along the direction of propagation of merger shocks. However, clusters hosting double relics are known to be perturbed. Observations provide evidence that radio relics are aligned with the merger axis of clusters \citep[][]{2011A&A...533A..35V} and cosmological simulations show that the gas density along 
the major axis of merging clusters is significantly higher than the average profile, up to $\sim 20-30$ \% close to the virial radius \citep[][]{va11scatter, 2013MNRAS.431..954K,va13clumping}. X-ray observations also suggest departures of this level in the outer parts of clusters \citep[][]{2012A&A...541A..57E,2013A&A...551A..22E,2014MNRAS.437.3939U,2015arXiv150104095M}. \\
We tested the impact of a systematically $20$ \% higher upstream gas density in all our clusters, producing the results given in Fig. \ref{fig:second} (left).  The enhanced density exacerbates the problems with the $\gamma$-ray emission because this scales as $\propto n^2$. The localised presence of denser clumps along the major axis of relics can only make this problem worse. We conclude that the hadronic emission predicted by our baseline model probably underestimates the level of $\gamma$-ray emission that DSA should produce in these objects.
Moreover, our assumption on the upstream gas density can be relaxed by considering that very likely before the heating by the crossing merger shocks the medium
in front of the relic had a temperature $\leq T_{\rm 500}$. As a very conservative case, we considered a pre-shock temperature lower by a factor $\sim 2$ compared to $T_{\rm 500}$, corresponding to a $M \approx 2$ shock. The right panel of Fig. \ref{fig:second} shows the result of this test. The predicted $\gamma$-ray emission is somewhat reduced compared to the "Bcap" model, most notably in the single-injection case because the propagation history of the single shock is the only crucial parameter. In this case the mean emission is a factor $\sim 2$ below the {\emph Fermi} limits.  However, in this case even more relics in all models require  $\sim 10 \mu G$ fields in more objects because a decrease in the upstream temperature in Eq.\ref{eq:hb} must be balanced by an increased magnetic field to match the radio power. 
In a more realistic case, we expect that the two above effects are combined since large-scale infall pattern along the major axis of merger clusters push cold dense un-virialized material further into the virial radius of the main halo, and therefore the problems of our DSA modeling of radio relics should become even worse.

\subsubsection{Radial dependence of the Mach number}
\label{subsubsec:Mrad}
\begin{figure*}
    \includegraphics[width=0.45\textwidth]{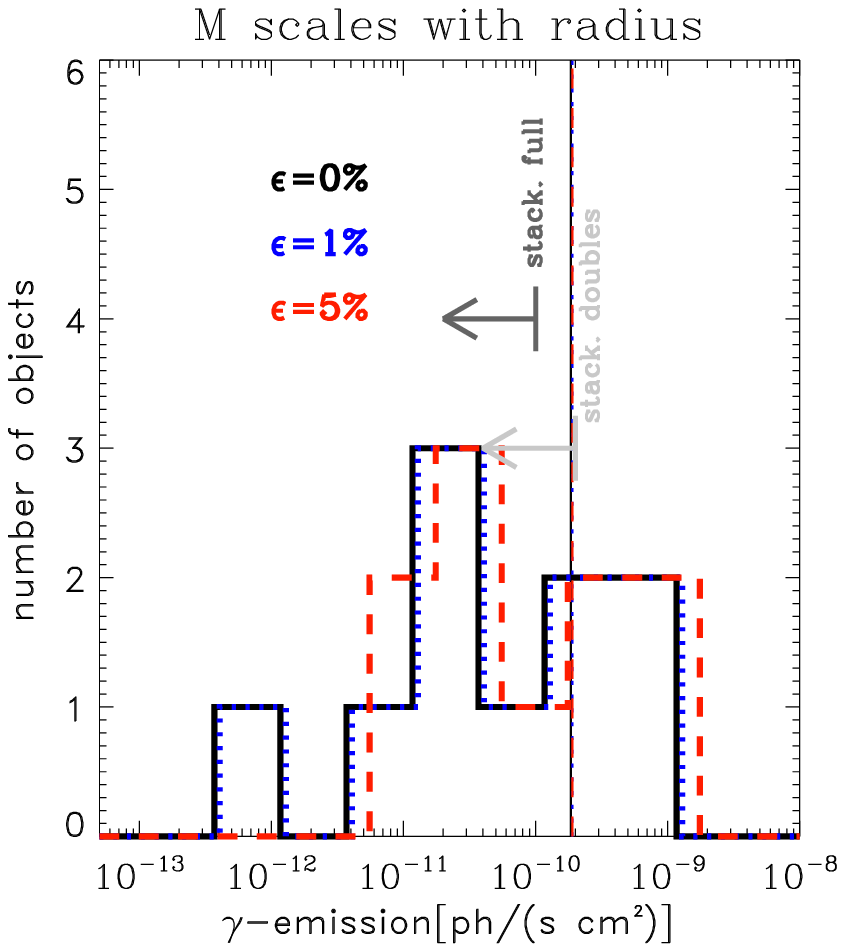}   
    \includegraphics[width=0.45\textwidth]{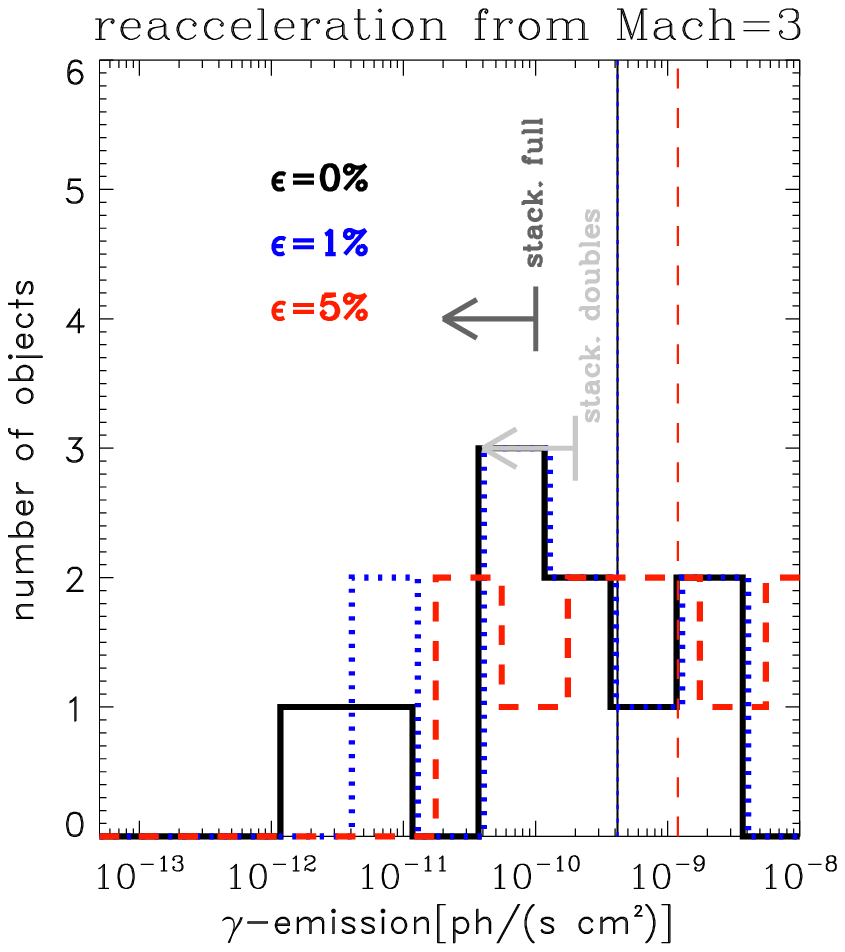}    
     \caption{Left: distribution of predicted $\gamma$-ray emission from our cluster sample, similar to Fig. \ref{fig:first} but here considering a radial scaling of the Mach number downstream of relics, $M(r) =M_{\rm 0} (r/R_{\rm relic})^{1/2}$. Right: same as in Fig. \ref{fig:first}, but here assuming that the re-acceleration in the $\epsilon=1$\% and $\epsilon=5$\% cases is done by a $M=3$ shock.} 
     \label{fig:third}
  \end{figure*}

Our assumption of a constant Mach number within the volume of interest follows from the assumption that the kinetic energy flux across shocks is conserved during their propagation. This implies that the dissipation of kinetic energy by these shocks is negligible, which is appropriate for the weak shocks considered here. Also the upstream medium is assumed to be isothermal (i.e. $n_{\rm u}(r) (M c_s)^3 S(r)/2={\rm constant}$ which gives $M \propto [n_{\rm u}(r) \cdot S(r)]^{-1/3} \approx \rm constant$ because $n_{\rm u}(r) \propto r^{-2}$ outside of the cluster core in the $\beta$-model). We also tested the possibility of a shallow radial dependence of the Mach number,  $M(r) =M_{\rm 0} (r/R_{\rm relic})^{1/2}$ (where $M_{\rm 0}$ is the Mach number estimated at the location of the relic), as in\citet{va14relics}. This was derived from the observed radial dependence of the radio spectral index of relics with radius \citep[][]{vw09}. However, when only giant radio relics are considered, this trend is not significant 
 \citep[][]{2012MNRAS.426...40B,fdg14}. To the best of our knowledge the average radial trend of Mach number in clusters was discussed only by \citet[][]{va09shocks,va10kp} and more recently by \citet{2014ApJ...785..133H}. All these works confirm a very shallow functional dependence with radius of the average Mach number of shocks, typically going from $M \sim 1.5-2$ in the centre to $M \sim 3$ in the cluster 
periphery. This trend is consistent with  $M \propto M_{\rm 0} (r/r_{\rm c})^{1/2}$ (with $M_{\rm 0} \sim 2$). In Fig. \ref{fig:third} we show the results if we impose a  $M \propto r^{1/2}$ instead of a constant Mach number. 
The predicted $\gamma$-ray emission downstream of relics is significantly reduced only in the single injection case ($\epsilon=0$), but the average emission of the sample still remains larger than both stacking limits. We conclude that the radial trend of Mach number with radius, as long as this is shallow as suggested by simulations, is not a crucial point. 
In the shock re-acceleration case, we also tested a scenario in which the re-acceleration is done by a $M=3$ shock instead of the $M=2$ as in our baseline model (Fig. \ref{fig:third}, right). In this case the requirement on the magnetic field are lowered and the number of relics for which we require $B_{\rm relic} \geq 10 \mu G$ is limited to one object (not shown). However, the predicted level of hadronic emission is now much increased and also the $\epsilon=1$\% run hits the \emph{Fermi} stacking limits for this sample of objects, while the $\epsilon=5$\% now predicts an average emission which is $\sim 4-5$ times larger than this.

\subsubsection{Mach number from the radio spectrum}
\label{subsubsec:Mach_MC}

\begin{figure}
    \includegraphics[width=0.45\textwidth]{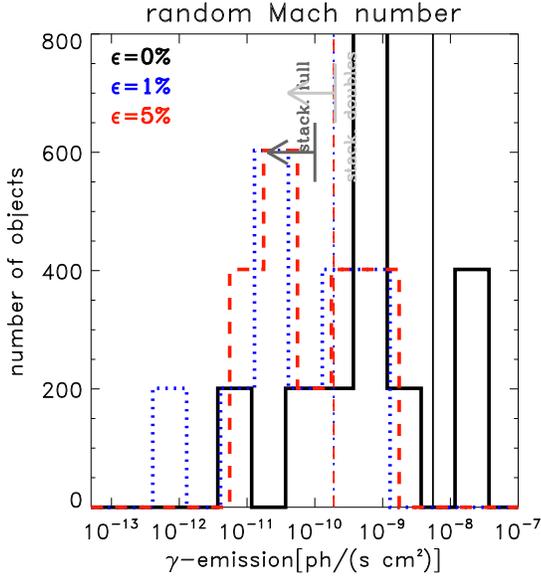}   
     \caption{Distribution of predicted $\gamma$-ray emission from our cluster sample including a random deviate from the Mach number derived from the radio, in the range $M=M_{\rm radio} \pm 0.5 M_{\rm radio}$ (see Sec.~\ref{subsubsec:Mach_MC} for details). We extracted 200 random deviates for each object and computed the downstream hadronic emission for the three reacceleration models. Differently from the previous Figures, in this case the vertical lines show the mean emission from the 200 stacked samples (i.e. we computed the mean emission within the sample, one time for each random extraction, and then computed the mean emission over the 200 realisations).} 
     \label{fig:Mach_MC}
  \end{figure}

The estimate of the  Mach number at the position of relics is crucial in our modeling, as it determines the level of CR acceleration in the DSA scenario we are testing. However, several effects can make the Mach number we derive from the radio spectrum in Sec.~\ref{subsec:algo} uncertain. 
In several observations the Mach number derived from the radio spectrum is found to be higher than the one estimated through X-ray analysis \citep[][]{2013MNRAS.433..812O,2013PASJ...65...16A}. The surface of complex shocks is described by a range of values of Mach numbers rather than by a single value \citep{sk13} and the radio emission will be dominated by electrons probing larger Mach numbers compared to the mean  \citep{2014ApJ...785..133H}. Additionally, radio observations with only a few beams across the relic can only produce integrated radio spectra, which are expected to be by $\approx 0.5$ steeper than the injection spectrum. A blend of several populations of electrons seen in projection can yield spectra with time-dependent biases, as recently discussed by \citet{2014arXiv1411.7513K}.
To assess this effect, we run a set of Monte Carlo methods and extract uniform random deviates within $M \pm \Delta M$, where $\Delta M \leq 0.5 M$. For each relic we randomly extracted 200 values of $\Delta M$ and compute the downstream $\gamma$-ray emission in all cases. 
Figure \ref{fig:Mach_MC} shows the results for the single injection and reacceleration models (colored histograms). In this case the simulated mean emission in the plot is the average of each set of 200 realisations (i.e. we first compute the mean emission for the cluster sample, for one random combination of extractions for $\Delta M$, and then compute the average emission and dispersion within the full dataset of 200 random realisation). The number of 200 realisation was chosen based on the fact that the errors in the mean emission do not change significantly for larger numbers of realisations. 
Compared to our fiducial model, a random variation in the assumed Mach number overall increases the level of hadronic $\gamma$-ray emission, suggesting that on average our baseline model underestimates the total emission. The underestimate is obviously more significant in the single-shock model ($\epsilon=0$), where the mean emission is $\sim 5$ smaller than what we obtain as an average from the 200 random extractions. In the recceleration cases, the effect is smaller, and the fiducial model probably underestimates the hadronic emission by a factor $\sim 1.5-2$. For the same set of random extractions, the problem with requiring too large magnetic fields for a significant fraction of objects ($\sim 1/3$ in the $\epsilon=0$ and $\epsilon=1$ \%) remains (not shown).
We conclude that realistic uncertainties in the Mach number do not change the robustness of our results.

\subsubsection{Viewing angle}

\begin{figure}
    \includegraphics[width=0.45\textwidth]{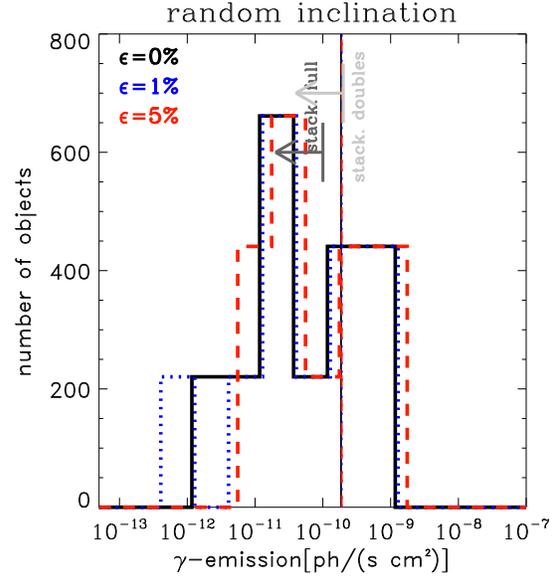}   
     \caption{Similar to Fig.\ref{fig:Mach_MC}, but here by extracting 200 random values for the angle in the plane of the sky for each
     relic, $|\delta \omega| \leq 30$ degrees.}
     \label{fig:angle_MC}
  \end{figure}

So far we assumed that all relics trace shocks which propagate exactly in a plane perpendicular to the line of sight. Numerical simulations
of relics support this scenario and limit the inclination along the line of sight of the propagation plane down to $\leq 10-20$ degrees \citep[][]{vw12sim,ka12}. In general, simulated radio relics assuming DSA resemble the observational properties of most relics only when they lie close to the plane of the sky  \citep[][]{va12relic,sk13}.
However, very small inclinations can be present and we checked if the inclusion of small ($|\Delta \omega| \leq 30$ degrees) along the line of sight can alter the picture in any significant way. Similar to the previous test, we randomly extracted 200 values of $\Delta \omega$ for a uniform distribution in the "Bcap" model, and accordingly recalculated the total volume spanned by shocks (which can only become {\it bigger} compared to the $\Delta \omega=0$ case), and computed the average value of the 200 realisations of cluster stackings. 
Fig. \ref{fig:angle_MC} shows the results of this test. The average $\gamma$-ray emission from all realisations is of the order the fiducial model ($\Delta \omega=0$) and at the level of the {\emph Fermi} stacking for this sample, and larger than the stacking by \citet{fermi14}. The outcome in the distribution of magnetic fields at relics is even worse than in the fiducial case, because in the case of large angles along the line of sight the relics are located further out, where the gas density is lower than in the $\Delta \omega=0$ case, and the magnetic field
must increase dramatically to match the radio power. As for all previous tests, we conclude that the presence of small but unavoidable projection effects has a small effect. However, on average these projection effects should yield an even larger hadronic emission from our dataset if DSA is at work.

 \subsubsection{Uncertainties in Cosmic Ray physics}
 
The  efficiencies that we have tested produce the lowest amount of CR-acceleration \citep[][]{kr13}. However, previously suggested functions for the $\eta(M)$ acceleration efficiency \citep[e.g.][]{kj07} give a larger injection of CRs from $M \leq 5$ shocks, and can only make the problem with \emph{Fermi} limits worse.
  Other uncertainties in the physics of CRs after their injection are briefly discussed here.
Outside of cluster cores, the CRs are only weakly subject to hadronic and Coulomb losses, owing to the low gas density.  For the sake of our analysis, it does not make any difference if they diffuse in the cluster volume at constant radius  (since their contribution to the $\gamma$-ray emission only depends on the gas density and not on the exact location in the cluster atmosphere). Only diffusion in the vertical sense can change our estimate. However, this cannot be a big effect since CRs are thought to be frozen into tangled magnetic fields, and in this case their spatial diffusion is slow (i.e. $\tau \sim 2 \cdot 10^8 \rm yr  ~(R/Mpc)^2  ~(E/GeV)^{-1/3}$ for a constant $B=1 ~ \rm  \mu G$ magnetic field and assuming Bohm diffusion, e.g. \citealt{bbp97}). 
More recently, it has been suggested that the fast ($v_{\rm streaming} \geq v_{\rm A}$, where $v_{\rm A}$ is the Alfv\'{e}n velocity) streaming of CRs can progressively deplete the downstream region of the shock and reduce radio and 
hadronic emission \citep[e.g.][]{2011A&A...527A..99E}, offering a way to reconcile hadronic models for radio halos with the observed bimodality in the distribution of diffuse radio emission in clusters \citep[][]{gb07}. However, the validity of this scenario is controversial \citep[e.g.][]{2013arXiv1306.3101D,2013MNRAS.434.2209W} and this mechanism has been suggested to be maximally efficient in relaxed clusters. Instead, our sample of clusters with double relics is all made by objects with a very unrelaxed X-ray morphology. In the case of turbulent intracluster medium, the detailed calculation by \citet{2013MNRAS.434.2209W} shows that even fast streaming can also rapidly diminishes the $\gamma$-ray luminosities
in the $E=300-1000$ GeV energies probed by imaging air Cerenkov telescopes (\emph{MAGIC, HESS, VERITAS}), but not in the lower energies  probed by \emph{Fermi}. Therefore, the latter is a more robust probe of the CR injection history, and our modeling here is well justified. 
Finally, our work neglects the (re)acceleration of CRs by other mechanisms, such as turbulent reacceleration in the downstream of relics \citep[e.g.][]{2004MNRAS.350.1174B}, which should re-energise radio emitting electrons as well CR-protons. Including this effect in our simple modeling is beyond the goal of this work, however its net effect can only be that of further increasing the mean emission we predict here.

\subsection{Other observational proxies: inverse Compton emission}

Relativistic electrons accelerated by shocks can also emit in the hard-X ray band through Inverse Compton (IC) emission  \citep{1979ApJ...227..364R,1998ApJ...494L.177S}. In principle, this can offer a complementary way of testing our models without having to make assumptions for the magnetic field.
Hence, we have computed the IC emission from each object in our sample, under the assumption of stationary shock acceleration following \citet{1999ApJ...520..529S}:
\begin{equation}
\epsilon_{\rm IC} \approx 0.17 \frac{E_{\rm CR,e}}{\Delta t} (\gamma \leq 5 \cdot 10^3),
\end{equation}
where $\Delta t$ is given by the shock
crossing time for each radial shell, $E_{\rm CR,e}$ is the energy of CR-electrons injected by the shock, and our integration is limited to the cooling region close to each shock. This is computed using Eq. (12) in \citet[][]{ka12}:

\begin{equation}
l_{\rm cool} \approx 890 \rm ~kpc \frac{v_{\rm s}}{10^3 \rm km/s} \cdot \frac{B^{1/2}}{B^{2}+B_{\rm CMB}^{2}} \cdot \frac{\nu}{1 \rm ~GHz}(1+z)^{-1/2} ,
\end{equation}
where $z$ is the redshift.
Here we show our prediction for the last investigated case, where we cap the magnetic field at $B=10 ~\rm \mu G$ and allow for the inclusion of CRs re-acceleration when necessary to match the observed radio power (Sec.~ref{subsubsec:Bfield}).

\begin{figure}
    \includegraphics[width=0.45\textwidth]{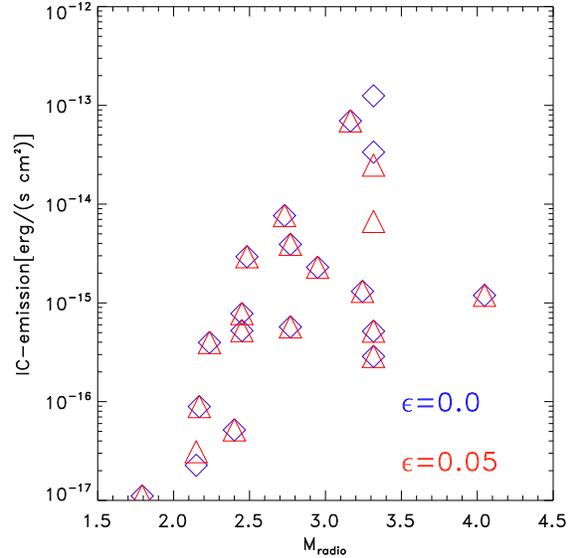}   
     \caption{Forecast of Inverse Compton emission from our simulated relics, in the extreme cases of $\epsilon=0$ (blue) and $\epsilon=0.05$ (red).}
  \label{fig:IC}
\end{figure}

Figure \ref{fig:IC} gives our predicted emission for the single injection case (blue) and for the extreme re-acceleration case ($\epsilon=0.05$), to emphasize
the weak dependence of the predicted IC emission on the assumed re-acceleration model. Table 3 gives the predicted flux in IC emission from the downstream of all double relics in the  $[20-100] \rm keV$ range, and the assumed radiative lengths in the downstream region.
In all cases the predicted emission lies below the detection threshold by the hard-X ray satellite \emph{NUSTAR}{\footnote {http:\/\/www.nustar.caltech.edu\/ }}, which has been estimated to be of the order of a few $\sim 10^{-12} \rm erg/(s  ~cm^2)$ in the case of the recent observations of the Coma cluster  \citep{2014arXiv1411.1573G} and of the Bullet cluster \citep{2014ApJ...792...48W}. However, a significantly lower sensitivity might be reached in the case of peripheral relics, given that in the latter clusters the contamination from the hot thermal gas in the hard-X ray range hampers the detection of the IC signal. 
This might be the case for the most powerful targets in our sample,  represent by A3376 (both relics) and by the most powerful relic in ZwCLJ2341. In these cases, the large distance from the centre of host clusters ($\sim 1.2-1.3$ Mpc) might indeed offer a better chance of detection of the IC signal. In the next years, the \emph{Astro-H} satellite should be able to probe the inverse Compton emission in the same clusters \citep[][]{2014SPIE.9144E..26A,2015arXiv150106940B}.

\begin{table*}
\label{tab:tab_IC}
\caption{Forecast of IC emission from the downstream region of relics in our simulated clusters, for $[20-100]$ keV. The predictions are here only given for the $\epsilon=0$  using our model with a capping of the magnetic field at $10  ~\rm \mu G$, see Sec. \ref{subsubsec:Bfield} for details), as all others yield extremely similar results.}
\centering \tabcolsep 2pt 
\begin{tabular}{c|c|c|c|c|c|c|c|c}
object & $M_1$ & $M_2$ & $\log_{\rm 10}(e_{\rm IC,1})$ & $\log_{\rm 10}(e_{\rm IC,2})$ & $l_{\rm cool,1}$ & $l_{\rm cool,2}$ & $B_1$ & $B_2$\\ 
   &  & & $[{\rm erg/(s ~cm^2)}]$ & $[{\rm erg/(s~cm^2)}]$ & [kpc] & [kpc] & [$\mu$G] & [$\mu$G] \\ \hline
A3376 &  3.3 &  3.3 & -13.98 & -13.19& 273.6 &  191.3 & 2.25 & 0.54\\
A3365 &  2.1 &  1.8 & -16.78 & -18.78&  87.6 &   48.6 & 5.41 & 8.69\\
A1240 &  3.3 &  2.8 & -15.29 & -15.24& 100.7 &   96.5 & 0.99 & 0.89\\
A2345 &  2.8 &  2.2 & -14.41 & -15.40& 107.2 &  131.1 & 0.60 & 0.97\\
RXCJ1314 &  2.4 &  2.4 & -15.11 & -15.28&  86.4 &  111.2 & 0.38 & 0.65\\
MACSJ1149 &  3.3 &  2.4 & -15.54 & -16.29& 101.7 &   96.4 & 1.75 & 1.46\\
MACSJ1752 &  3.2 &  4.0 & -14.88 & -14.92& 122.2 &   89.1 & 3.21 & 5.92\\
A3667 &  1.7 &  3.8 & -14.53 & -14.12& 164.4 &  159.7 & 1.56 & 1.32\\
ZwCLJ2341 &  1.8 &  3.2 & -16.96 & -13.16& 185.7 &   97.7 & 2.15 & 0.36\\
PLCKG287 &  2.9 &  2.2 & -14.64 & -16.05& 125.4 &  128.4 & 1.40 & 3.89\\

\end{tabular}
\end{table*}

\section{Discussion and conclusions}
\label{sec:conclusions}
Most of the evidence from X-ray and radio observations suggests a link between radio relics and merger shocks: merger axes of clusters and relic orientations correlate \citep[][]{2011A&A...533A..35V}, the power of radio relics scales with X-ray luminosities \citep[][]{2012MNRAS.426...40B,fe12} and mass \citep[][]{fdg14} of the host cluster.
Cosmological simulations produce emission patterns consistent with observed radio relics just using a tiny fraction of the kinetic
energy flux across shock waves \citep[e.g.][]{ho08,2008MNRAS.385.1242P,2009MNRAS.393.1073B,sk11,va12relic,2012MNRAS.420.2006N,sk13}.\\

Still, a number of recent observations have revealed some open issues, including uncertain merger scenarios \citep[e.g.][]{2013MNRAS.433..812O}, departures from power-law spectra \citep[e.g.][]{2014MNRAS.445.1213S,2014arXiv1411.1113T}, missing associations between radio emission and X-ray maps \citep[e.g.][]{2011MNRAS.417L...1R,2014MNRAS.443.2463O}, efficiencies problems \citep[e.g.][]{2011ApJ...728...82M,2012MNRAS.426...40B}, inconsistencies between Mach numbers derived from X-ray and radio observations \citep[e.g.][]{2012MNRAS.426.1204K,2013PASJ...65...16A,2013MNRAS.433.1701O} and apparent connections to radio galaxies \citep[][]{2014ApJ...785....1B}.\\

In this work, we used a simple semi-analytical model of expanding merger shocks in clusters to reconstruct the propagation history of shocks leading. We used a spherically symmetric model  and assumed that cosmic ray protons are trapped in the intracluster medium on all relevant timescales. A range of realistic scenarios for the acceleration of relativistic electrons and protons via DSA, varying the upstream gas conditions, the shock parameters and the budget of pre-existing cosmic rays, gives very similar results. In all realistic scenarios, a significant fraction of our objects ($\sim 1/2-1/3$)  has difficulties in matching at the same time the observed radio emission and the  constraints imposed by the \emph{Fermi} limits, unless the magnetic field in all problematic objects is much larger than what usually considered realistic ($\gg 10 ~\rm \mu G$).
The scenario in which radio emitting electrons comes from the re-acceleration of pre-existing electrons  \citep[][]{ka12,pinzke13} can alleviate the tension with \emph{Fermi} if the pre-existing electrons are not the result of previous injection by shocks as we investigated here, but are instead released by mechanisms that mostly inject leptons (e.g. leptonic-dominated jets from AGN), as already discussed in \citet{va14relics}.

Based on our semi-analytical model, the standard DSA scenario with thermal leakage that predicts that $E_{\rm CRe} \ll E_{\rm CRp}$ cannot simultaneously explain radio relics and produce less $\gamma$-radiation than the upper limits from \emph{Fermi}, unless unrealistically large magnetic fields are assumed at the position of relics (e.g. $B_{\rm relic} \ge 10-100  ~\rm \mu G$).
This result is very robust, at least in the statistical sense, against all investigated variations of our fiducial parameters for the modeling of the shock acceleration of CRs. Additional effects that go beyond our idealized modeling of cluster mergers, e.g. a clumpy ICM, a succession of mergers and the additional acceleration of CRs by AGN, supernovae, turbulence or reconnection exacerbate this discrepancy. \\

Despite its obvious degree of simplification, a semi-analytical method is useful to tackle the case of double relic systems. In these systems it is reasonable to assume that most of the energetics  is related to the observed pair of giant merger shocks. Their shape and location is rather regular and symmetric with respect to the cluster centre, suggesting that one can make reasonable estimates for their propagation history. This setup allows us to run very fast testing of different possible acceleration scenarios, and as we showed in our various tests it generally gives a {\it lower limit} on the expected $\gamma$-ray emission. This method is meant to be complementary to fully cosmological numerical simulations where the effects of multiple shocks, particle advection and cooling, as well as inputs from galaxy formation and other mechanisms can be taken into account at run-time \citep[e.g.][]{pf07,scienzo}. However, a thorough exploration of models is computationally demanding because of the required high resolution and the complexity of the numerics. Also the agreement between different numerical techniques on this topic is still unsatisfactory \citep[see discussion in][]{va11comparison}.

\begin{figure}
    \includegraphics[width=0.495\textwidth]{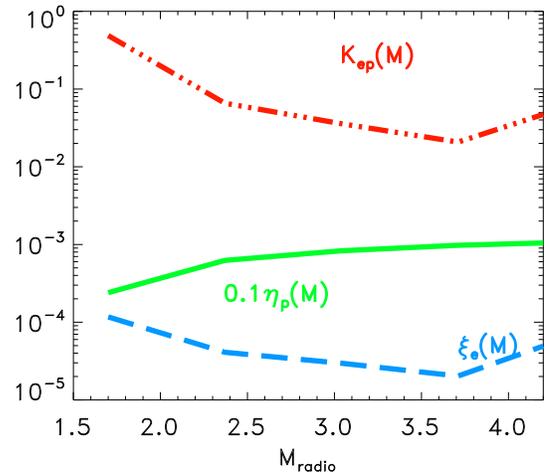}   
     \caption{Acceleration efficiency of CR-protons (green, rescaled by a factor $\times 10$ down) and CR-electrons (blue), and electron to proton acceleration ratio (red) allowed by our combined radio and $\gamma$-ray comparison with observations. In this case, we assumed a fixed magnetic field of $B=2 \mu G$ for all relics.}
  \label{fig:last}
\end{figure}

Another way of illustrating our result is found by rescaling the efficiency for proton acceleration, $\eta(M)$, such that the upper limits from \emph{Fermi} are not violated. In the case without pre-existing CRs, this is a simple exercise as we only need to rescale $\eta(M)$ for each relic separately, and compute the average of the efficiencies for each bin of the Mach number (here we chose a bin size of $\Delta M=0.6$ to achieve a reasonable sampling of the sparse distribution of Mach numbers in the dataset). Here, we keep the magnetic field fixed at $B=2 \mu$G as suggested by recent observations \citep[][]{fdg14}. The result is shown in Fig.~\ref{fig:last}, where we show the maximally allowed acceleration efficiency for CR-electrons, protons, as well as $K_{\rm e/p}$ as a function of $M$. This relation results from a somewhat coarse simplification of the problem but it is a rough estimate of the acceleration efficiencies in weak ICM shocks. 

For shocks with $M \leq 2$, the flux ratio of injected electrons is larger than that in protons, $K_{\rm e/p} \sim 1-100$, at odds with standard DSA (even including re-accelerated electrons).  For $M \geq 2.5$ the acceleration efficiency of protons can become significant ($\sim 10^{-3}-10^{-2}$) while the acceleration efficiency of electrons flattens and $K_{\rm e/p} \sim 10^{-2}$. The functional shape of the acceleration efficiency for protons is consistent with the \citep[][]{kr13} model, but the absolute normalisation is lower by a factor $\sim 10-100$. 

A possible solution has been suggested by \citet{2014ApJ...788..142K}, who assumed that electrons and protons follow a $\kappa$-distribution near the shock transition. A $\kappa$-distribution is characterized by a power-law rather than by an exponential cutoff at high energies, thus ensuring a more efficient
injection of high-energy particles into the DSA cycle. This distribution is motivated by spacecraft measurements of the solar wind as well as by observations of HII and planetary nebulae \citep[e.g.][and references therein]{2012ASSP...33...97L}.  \citet{2014ApJ...788..142K} explored the application of the $\kappa$-distribution to $M \leq 2$ shocks in the ICM, and concluded that the distribution can have a different high-energy tail as a function of the shock obliquity and of the plasma parameters. In the ICM, the distribution might be more extended towards high energies for electrons than for protons thus justifying a higher acceleration efficiency for electrons than for protons. However, in order to explain the origin of these wider distributions, one must resort to detailed micro-physical simulations of collisionless shocks.\\

The most promising explanation for the non-observation of $\gamma$-rays has been suggested by \citet[][]{2014arXiv1409.7393G} who studied the acceleration of electrons with particle-in-cell (PIC) simulations under conditions relevant to merger shocks. They showed that $M \leq 3$ shocks can be efficient accelerators of electrons in a Fermi-like process, where electrons gain energy via shock drift acceleration (SDA). The electron gain energy from the motion of electric field and scatter off oblique magnetic waves that are self-generated via the firehose instability. They found that this mechanism can work for high plasma betas and for nearly all magnetic field obliquities. However, these simulations have been performed in 2D, and could not follow the acceleration of electrons beyond a supra-thermal energy because of computing limitations. At the same time, hybrid simulations of proton acceleration by \citet{2014ApJ...783...91C} have shown that the acceleration efficiency is a strong function of the obliquity 
angle. If indeed the magnetic field in radio relics is predominantly perpendicular to the shock normal, as found e.g. in the relic in the cluster CIZA 2242.2+5301, then the prediction is that the acceleration efficiency of protons is strongly suppressed, thus explaining the non-detection of hadronic emission. It remains to be seen if the results of these simulations hold in 3D, with realistic mass ratios between electrons and protons and coupled to a large scale MHD flow. It is also not clear whether the magnetic field is quasi-perpendicular in all the relics of this sample and how the alignment of the magnetic fields with the shock surface observed on large scales can be scaled down to scales of the ion gyro radius.

\section*{acknowledgments}

FV and MB acknowledge support from the grant FOR1254 from the Deutsche Forschungsgemeinschaft. DE and BH thank Andrea Tramacere and Christian Farnier for their help with the development of the Fermi tools. We acknowledge fruitful scientific discussions with F. Zandanel, A. Bonafede, F. Gastaldello and T. Jones for this work.

\bibliographystyle{mnras}
\bibliography{franco}

\appendix

\end{document}